\documentclass[aps,article,twocolumn,superscriptaddress]{revtex4}
\usepackage{graphics} 
\usepackage{amsmath}
\usepackage{color}
\DeclareMathOperator{\sech}{sech}

\begin{document}
\title{Ultra slow electron holes in collisionless plasmas: stability at high ion temperature}
\author{Debraj Mandal}
\affiliation{Aix-Marseille University, CNRS, Marseille, 13397, France}
\author{Devendra Sharma}
\affiliation{Institute for Plasma Research, HBNI, Bhat, Gandhinagar, India, 382428}
\author{Hans Schamel}
\affiliation{Physikalisches Institut, Universit\"{a}t Bayreuth, D-95440 Bayreuth, Germany}

\date{\today}

\begin{abstract}
Numerical simulations recover ultra slow electron holes (EH) of 
electron-acoustic genre
propagating stably well below the ion acoustic speed where the ion response
disallows any known pure electron perturbation. 
The reason of stability of EH at high ion temperature ($T_{i}> T_{e}$) is 
traced to the loss of neutralizing cold ion response. 
In a background of cold ions, $\theta=T_e/T_i\gg 1$, they have an ion 
compression that accelerates to 
jump over a forbidden velocity gap and settle on the high velocity
tail of the electron distribution $f_e$, confirming to a recently identified
limit of the nonlinear dispersion relation. 
For $\theta=T_e/T_i \le 1$, however, 
the warm ions begin to supplement the electron response
transforming the ion compression to decompression at the hole location and
triggering multiplicity of the scales in trapped electron 
population which prompts an immediate generalization of the basic EH theory.
\end{abstract}



\maketitle
\section{Introduction \label{introduction}}
The collective excitations in collisional plasmas are well represented by discrete
linear waves below the amplitudes where the convective nonlinearity 
of fluid formulation begins to assume significance. In hot collisionless 
plasmas, however, the earliest (often vanishing) threshold to nonlinear 
behavior is introduced by the kinetic effects 
such that the
waves very fast achieve coherency at unusually low amplitudes.
The first accessible class of nonlinear collective 
excitations in hot plasmas therefore, in practice and in most numerical 
simulations, is that of the nonlinear particle trapping equilibria, 
such as the non-isothermal ion acoustic solitary waves,  solitary and cnoidal 
electron and ion holes or various forms of double layers
\cite{Saeki79, S79, schamel:72, S86,krall:na,bernstein:i,dupree83:t}. 
These nonlinear modifications often render stability to rather exotic 
excitations in the plasma,
for example, electron acoustic perturbation slower than ion acoustic speeds
\cite{johnston:t,lesure:m1,petkaki:p,mandal:d1} and electron holes structures
in the circular particle beam, or synchrotron, experiments 
\cite{S97,BWLS04,LS05}. 

%
The {\em simplest} nonlinear
analytic approach to the
experimentally and numerically observable class of excitations
\cite{mandal:d,osmane:a,eliasson:b1,saeki:k1,pickett:j2,osborne:a2,gurevich:a,wilson:l}
works by invoking a fixed ionic background and a thermal Maxwellian 
approximation for equilibrium electron distribution
in the Vlasov analysis, for example in all well known linear 
\cite{landau,vancampan} and nonlinear \cite{schamel:h1} approaches.  
For the trapped electrons, however, a 
variety of Ans$\ddot a$tze is in principle 
possible but using a, thermalized (single parameter), distribution of 
trapped particles produces the simplest class of nonlinear solutions. 
Additionally, as long as the equilibrium distributions are thermalized 
and identical to 
those used for obtaining linear modes, the nonlinear solutions can still 
be identified as corresponding to the well known linear modes, however having 
noticeable (nonlinear) modification by trapped particles.
%
For an easy reference, this limit of nonlinear Vlasov treatment is termed here 
as a Special Limit of Correspondence (SLC) of the general nonlinear Vlasov 
framework since the computer simulations of structures in collisionless 
plasmas are best interpreted by an approach in SLC, given the unavoidable 
numerical thermalization effects in them. We, for example, apply
the one developed extensively by Schamel 
and co-authors \cite{schamel:h1} which introduces amplitude dependence 
in the dispersion and removes much of discreteness of the linear wave
solution space.

It was recently discovered \cite{mandal:dnjp}, that 
the discreteness of the linear modes (distinct roots of 
linear dispersion function) also exists in the nonlinear solutions space
(as corresponding band gaps) which was originally 
understood to be a continuum \cite{schamel:h1} solution space. 
These band gaps, or forbidden velocity ranges, were identified to be allowed 
also by the hole theory after the simulations in \cite{mandal:dnjp} could 
achieve no stably propagating EH structures in particular velocity ranges. 
In more specific terms, the nonlinear EH structures were 
noted as unstable (i.e., not propagating coherently but accelerating) below 
a critical velocity value
which almost ruled out existence of any electron holes slower than nearly 
the ion acoustic speed, explaining several past observations of accelerating 
holes in the simulations \cite{zhou:c,eliasson:b,schamel:hms}. 
By applying the
special correspondence, of the nonlinearly obtained limiting velocity value 
with the ion acoustic phase 
velocity in the linear theory, it appeared that the acoustic structures would 
not propagate below
the ion acoustic speed in a typical plasmas where $T_{e}$ is sufficiently 
larger than $T_{i}$, 
essentially because of possibility of neutralizing ion response below 
this velocity (slow enough time scale).
In this paper we however present a set of simulations showing that ultra 
slow electron holes regain their stability at large enough ion temperature which 
exceeds electron temperature. 
The associated nonlinear analytics shows that the band gap is a dynamical one 
and may indeed be buried with the changing ion temperature, 
showing no 
minimum
cutoff velocity (e.g., the ion acoustic speed) for structures with no ion 
trapping. 

With no significant contribution of resonant ions
and a 
decompressed electron density, the observed ultra slow EH correspond to the
electron-acoustic structures. They however have an unusual ion density profile 
which is also decompressed, in contrast to the ion compression in their usual 
velocity regime. 
In the conclusion of this paper we finally highlight an important issue 
that the theoretical recovery of these slow electron hole structure 
under the basic EH theory may be possible only by a further 
extension of the theory. 
Although such an extension and greater details of this analytic 
aspect is addressed in a dedicated forthcoming article \cite{MSS19}, the idea 
mainly pertain to phase-space topology of the trapped electron 
density of the observed ultra-slow stable electron hole structures 
summarized in following statements. While the observed 
slow EH are recovered to have a dip-like trapped electron density, the lowest 
order electron hole theory prescribes them to have exclusively a humped 
structure. A modification of the lowest order electron hole theory is 
however possible by appropriate higher order corrections, allowing it to 
yield a dip-like trapped electron density structure, as recovered 
numerically for these ultra slow structures, without causing any 
characteristic change in the associated pseudo-potential structure.

This paper is organized as follows. We present the results of our high 
resolution Vlasov 
simulations in Sec.~\ref{simulation}. The analytical model following 
the SLC of the Vlasov formulation is discussed and used to describe the 
results in Sec.~\ref{model}. The discussion of the physics of the observation 
and requirement of appropriate extension of the EH theory is highlighted in 
Sec~\ref{discussion} and the summary and conclusions 
are presented in Sec.~\ref{conclusion}.
\section{Simulation results \label{simulation}}
We performed Vlasov simulations using the 
Flux-Balance \cite{fijalkow:e,mandal:dconf}
technique for both electrons and ions in the $x$-$v$ space with 
8192 $\times$ 16384 dual mesh grid. A well localized initial perturbation 
is used in the electron distribution function with the analytic form of 
perturbation,
%
\begin{eqnarray}
f_{1}(x,v)= - \epsilon~{\rm sech}\left[\frac{v-v_1}{L_{1}}\right]{\rm sech}^4[k(x-x_{1})] 
\label{perturbation}
\end{eqnarray}
%
where $\epsilon$ is the amplitude of the perturbation, $L_{1}$ is the
width of the perturbation in the velocity dimension and $k^{-1}$ is its 
spatial width. 
\begin{figure*}
\includegraphics{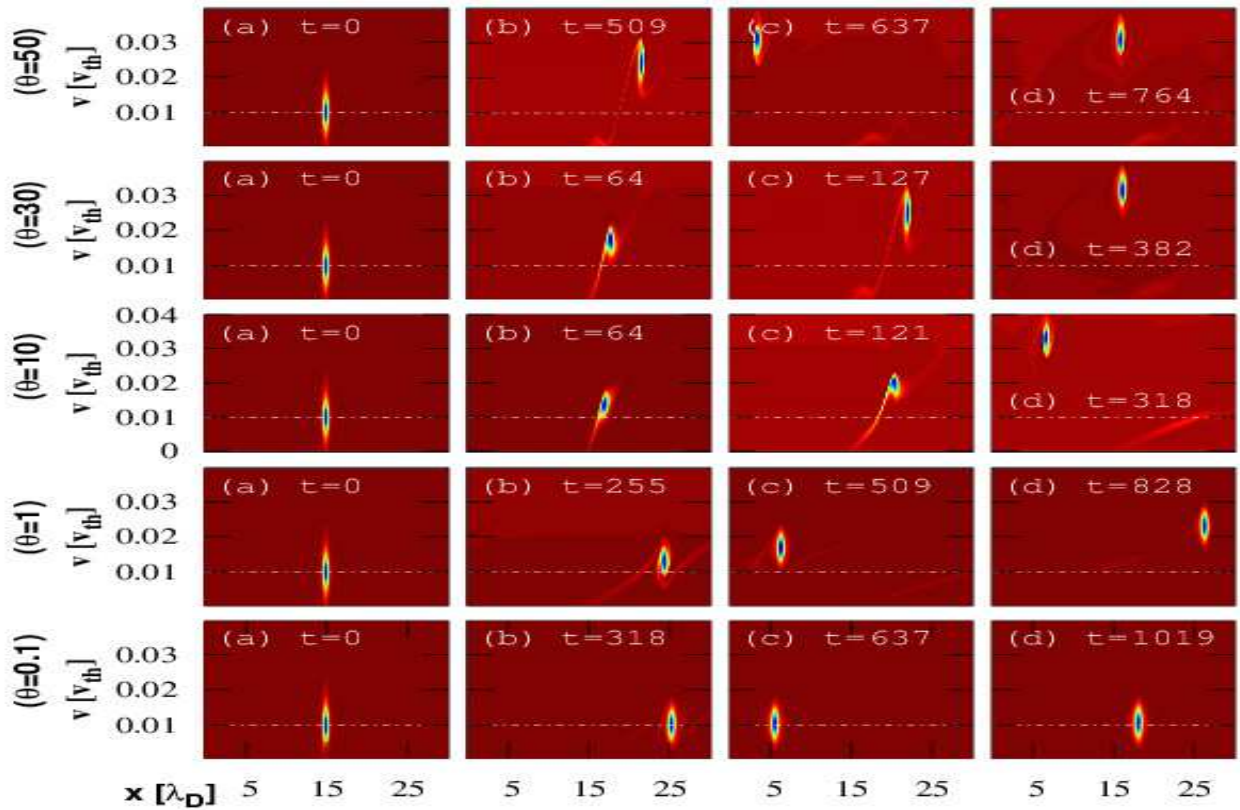}
\caption{Evolution of the electron phase-space perturbation in all the five 
cases. In case-5 ($\theta=0.1$) the electron hole is not accelerated from its
initial perturbation location. In all other cases they accelerate but their
evolution time is different. The cases with lower ion temperature 
(high $\theta$) take more time to construct a valid electron hole solution
from the initial perturbation. (The color scale is used for the value of 
electron distribution function, increasing linearly from blue to red.)
\label{phase_space_ev}}
\end{figure*}
%
The background equilibrium velocity distribution of the electrons is a 
shifted Maxwellian and that of the ions an unshifted  Maxwellian, given in 
normalized quantities by 
%
\begin{eqnarray}
f_{0e}(v)&=&\frac{1}{\sqrt{2 \pi}}\exp\left[-\frac{(v-v_D)^2}{2}\right]\\
f_{0i}(u)&=&\frac{F_0}{\sqrt{2 \pi}}\exp\left[-\frac{u^2}{2}\right]
\label{initial_f}
\end{eqnarray}
%
where $v$ is normalized by $v_{the}=\sqrt{\frac{T_e}{m_e}}$ and $u$ by 
$v_{thi}=\sqrt{\frac{T_i}{m_i}}$, respectively. Therefore 
$u=v \sqrt {\theta/\delta} $, where $\theta=T_e/T_i$ and $\delta=m_e/m_i$.
Subscripts $j=e,i$ correspond to electron and ion species, 
respectively. 
The factor $F_0$ is ratio of total ion and electron content in the 
simulation region ($\int\int f_{e}dx dv/\int\int f_{i}dx dv$), ensuring 
that the same number of ions and electrons are present in the simulation box.
In the simulation we use the Debye length 
$\lambda_{D}$, electron plasma frequency $\omega_{pe}$ and electron thermal 
velocity $v_{the}$ as normalizations for length, time and electron 
velocities, respectively. According to linear theory of plasma, 
the critical {\it linear threshold} ($v^*_D$), required minimum drift 
value $v_{D}$ for a current driven ion acoustic instability, becomes 
$v^*_D=0.053$ for $\theta=T_e/T_i = 10$ and $\delta=m_e/m_i = 1/1836$. 
For all the cases our choice of 
the drift velocity is $v_D=0.01$ which is well below the
linear threshold for those temperature ratios \cite{krall:na,landau:ld}.

We first present the evolution of the total electron distribution 
$f_{e}=f_{0e}+f_{1}$ in cases 1-5 having
$\theta=50$, $30, 10, 1$ and $0.1$, as plotted in Fig.\ref{phase_space_ev}, 
respectively, showing result of varying ion response in them.
Considering the temperature dependency, the ion acoustic wave
phase velocity in one dimension for these five cases are 
$C_{s}= 0.033, 0.036, 0.045, 0.093$, and $ 0.24 v_{the}$, 
where $C_{s}=(\sqrt{\delta}+\gamma \sqrt{\delta/\theta}) ~v_{the}$ 
with $\gamma=3$. Therefore in all the cases the initial electron
velocity perturbation location $v_{1}=0.01~v_{the}$ is well below the 
corresponding $C_{s}$ and also the drift velocity $v_{D}$ is well below 
the corresponding critical linear thresholds $v_{D}^{*}$. 
Moreover, in the last case the ion temperature is higher than the 
electron temperature. A nonlinear plasma response 
to the applied perturbation, in the form of amplitude dependent propagating 
coherent structures, is nevertheless seen in all the cases where
the perturbation of the form (\ref{perturbation}) is 
placed at $x=15 $ in the simulation box having the length
$L=30 $. The velocity perturbation location in all cases is $v_{1}=0.01$ with
phase-space widths of the perturbation $L_{1}=0.01$ along
the velocity dimension and $k^{-1}=10$ along the spatial 
dimension $x$. The strength of the perturbation is small: $\epsilon=0.02$.
As witnessed in our earlier simulations, being placed at such small velocity
the initial perturbation with $\theta>1$ is unstable and experiences an 
acceleration. 
For the last case with $\theta=0.1$, however, the time evolution 
of the contours of electron distribution function 
$f_{e}(x,v)$ in phase-space, presented in last row (from top) of 
Fig.~\ref{phase_space_ev}, shows that the perturbation is largely intact and, 
after a marginal readjustment of its $x$-$v$ space widths, continues its 
propagation with nearly the original velocity, $0.01 v_{the}$. 

\begin{figure}
\includegraphics{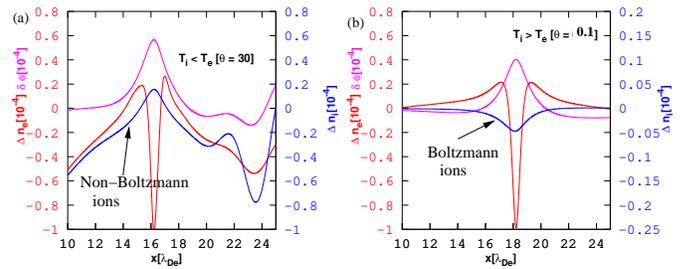}
\caption{Electron density, ion density and potential prfiles for
(a) $\theta$=30 and (b) $\theta$=0.1. 
\label{density-potential-profiles}}
\end{figure}

Considering insignificant contribution of resonant ions (a very narrow 
velocity range of ion trapping region, as compared to trapped electrons),
the stability of electron holes for small $\theta$ is once again 
understood to be detrmined by collective shielding effects rather than
resonant ion reflection \cite{dupree83:t}. 
In qualitative sense \cite{mandal:dnjp} 
it can be described as follows.
In a stable hole, the flux of the cold ion 
density expelled by the positive potential of the perturbation balances 
the flux of ions pushed in by the relative excess of hot electrons surrounding
the hole. Thus the stability is achieved at faster velocities because of 
smaller 
outflowing ion flux due to smaller exposure of background ions to the 
hole electric field \cite{mandal:dnjp}.
The stability at smaller velocity therefore presents an interesting case
and indicates a new mechanism underlying the stable holes to overcome 
destabilizing cold ion response that, in the usual case of colder ions, 
necessitates a minimum velocity for the 
stability. The slow holes observed in our simulation are found to achieve 
this stability by marginalizing the cold ion response in the limit 
$\theta\ll 1$.
We observe that the stability is achieved critically when the single 
(fully untrapped) ion population stops supplementing the response of cold 
electrons 
and instead begins to supplement the response of streaming Boltzmann 
electrons. This means the warm ions rather rarefact at the hole location in 
full accordance with the Boltzmann-like response of positive ions to a positive 
potential. 
This behavior of ion density is clearly visible in the ion density profiles
shown in Fig.~\ref{density-potential-profiles}(a) and (b) for large and small 
values of $\theta$, respectiely.

In the next section we examine this aspect more quantitatively and 
explain that these solutions are a
special class of Cnoidal Electron Holes representing the nonlinear 
solutions of the Vlasov equation.
\section{Analytical Vlasov model, gaps of existence and quasi-particle interpretation\label{model}}
The existence regimes of Cnoidal Electron Holes (CEHs) and their dependence on the ion temperature 
are now evaluated in more quantitative terms using the
non-perturbative nonlinear dispersion formulation of Schamel 
(see e.g. \cite{schamel:h1} 
and references therein). 
Note that the description below is limited to finding the thresholds that
bound the parameter regime in which the formal solutions of Vlasov equation, 
prescribed in \cite{schamel:h1}, represnt an undamped propagation.
A solution outside these thresholds does not satisfy nonlinear dispersin 
relation and therefore must undergo a transient, or phasemix, 
i.e. an evolution which is not covered by a nonlinear dispersion 
formulation that aimes to identify only the coherently propagating solutions.
A more detail description of the analytical model used here
is given in Appendix-A.  
The phase velocity of a settled vortex structure in electron phase space
is determined by the nonlinear dispersion relation (NDR) 
Eq.~(\ref{NDR_eq_Apdx}),
\begin{eqnarray}
k_{0}^2&-&\frac{1}{2}Z_{r}^{'}\left(\tilde{v}_{D}/\sqrt{2}\right)
-\frac{\theta}{2}Z_{r}^{'}\left(u_{0}/\sqrt{2}\right)=B,
\label{NDR}
\end{eqnarray}
where, $Z_{r}(x)$ is the real part of the plasma dispersion function. 
Depending on different values of $k_0$ and $B$ one gets different type of 
solutions, like solitary solution and cnoidal solutions. The right
hand side of the equation presents the contribution of free electrons and ions
and the left hand side presents the trapped particle contribution. 
Since in our case $v_{D}=0.01$, we can consider 
$ - \frac{1}{2}Z_{r}^{'}\left(\tilde v_D/\sqrt{2}\right) \sim 1$. For 
a solitary electron hole $k_0^2 = 0$, and the NDR 
Eq.~(\ref{NDR}) can be written as,
\begin{eqnarray}
 -\frac{1}{2}Z_{r}^{'}\left(u_{0}/\sqrt{2}\right)= \frac{1}{\theta}(B-1) =: D.
\label{NDR_reduced}
\end{eqnarray}
Therefore, for our conditions of no ion trapping the phase velocity of the 
solutions are
controlled by the ion temperature $\theta$ and electron trapping parameter 
$B(\beta,\psi)$.  

\begin{figure}
\includegraphics{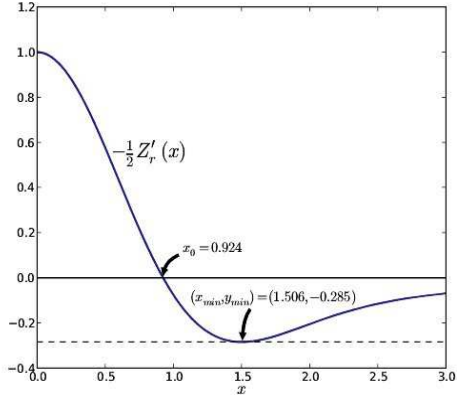}
\caption{Plot of $-\frac{1}{2} Z'_r(x)$ with $x$. 
\label{dispersion_fn}}
\end{figure}

 A solution of Eq.~(\ref{NDR_reduced}) together with $B>0$ decides 
quantitatively about the existence of solutions. The solubility demands, 
$-0.285 \le D \le1$, as Fig.~\ref{dispersion_fn} shows, in which 
$-\frac{1}{2} Z'_r(x)$ is plotted as a function of $x$. For $D \ge0$ 
(or $B\ge1$) one has one solution, and for $D<0$ (or $B<1$) one has two 
solutions.

%
There are accordingly three velocity regimes:
\begin{eqnarray}\nonumber
{\rm (i)}~~ 0.0 &\le& u_0 \le 1.307~ (=0.924 \sqrt2),  \qquad  1 \le B \le 1 + \theta\\
\nonumber
{\rm(ii)}~ 1.307 &<& u_0 \le 2.12~ (=1.5 \sqrt2), \quad  1-0.285~ \theta  \le B < 1\\
\nonumber
{\rm(iii)}~ 2.12 &<&u_0 < \infty,  \qquad \qquad \qquad 1-0.285~ \theta  \le B < 1.
\end{eqnarray}
The first two belong to the Slow Ion Acoustic branch (SIA), the third one to 
the ordinary Ion Acoustic branch (IA). In the second column the necessary 
conditions for $B$ are presented for given $\theta$, which are subject to 
$B>0$. This means that for the doublet of solutions, (ii), (iii), which satisfy 
the same constraints, ($-0.285 \le D < 0$), there is a division line for 
$\theta$ given by $\theta^*:=\frac{1}{0.285}=3.51$. For $\bf{cold}$  ions, 
$\theta > \theta^*$, any $B$ in $0<B < 1$  is admitted, whereas for 
$\bf{hot}$  ions, $\theta < \theta^*$, $B$ must satisfy 
$1-0.285~ \theta \le B < 1$.
In terms of $B$ we have therefore the following situation, for slow regime of
SIA given by (i) $1\le B \le 1+\theta$ there is no other choice for a 
solution. But for $1-0.285 \theta  \le B < 1$, the plasma has two choices for
establishing a solution, $u_0$ lies either on the faster part of the
SIA branch (ii) or on
the still faster velocity IA branch, regime (iii), with a gap in between. 

\begin{table}[h!]
\centering
\caption{$v_1$, $u_1$, $v_0$, $u_0$, $B$ and velocity regime}
\begin{tabular}{ |c|c|c|c|c|c|c| } 
 \hline
$\theta$ &$v_1$ &$u_1= \sqrt{\frac{m_i \theta}{m_e}}v_1$ &$v_0$ 
&$u_0= \sqrt{\frac{m_i \theta}{m_e}}v_0$ & $B$& v-regime\\ 
\hline
 0.1 & 0.01 & 0.136&0.01 &0.136&1.1& (i)\\ 
 1 & 0.01 &0.429 & 0.023&0.986&1.32& (i)\\ 
 10 & 0.01 &1.355 & 0.035&4.743&0.47& (ii) $\rightarrow$ (iii)\\ 
 \hline
\end{tabular}
\label{table:1}
\end{table}

Tab:~\ref{table:1} presents the initial perturbation velocity $v_1$ 
(in electron frame), $u_1$ (in ion frame), final velocity at 
settled state $v_0$ 
(in electron frame), $u_0$ (in ion frame), $B$ values (from 
Eq.~\ref{NDR_reduced}) and the velocity regime for these three cases 
$\theta = 0.1, 1$ and $10$. 
We define the case $\theta = 50, 30$ is identical with case $\theta =10$, 
because in all the three cases unstable electron holes saturate to same final
velocity $v_0 = 0.035 v_{the}$. 
Initially the perturbation is hence located for $\theta=(0.1,1)$ in (i) and for  $\theta=10$  in (ii). 
Therefore, in the cases-4 and 5 with $\theta = 1$ and $0.1$ the SEH will stay
in the same velocity regime (i), and for the case-3 with $\theta= 10$ there is 
a possibility of transition of solution from the velocity regime (ii) to (iii).
Since ion acoustic velocity in the ion frame is given by $c_s:= \sqrt \theta$
 we get the triplet $c_s= (0.32, 1, 3.16)$ and for the associated Mach numbers 
$M:=\frac{u_0}{c_s}$ the triplet (0.43, 1, 1.5). The SEH structure hence 
travels subsonically for hot and supersonically for cold ions, whereas it 
moves sonically for moderate ion temperatures. $\theta \approx 1$.
This furthermore implies from Eq.~(\ref{NDR_reduced}) $B: (1.1, 1.32, 0.47)$ or 
$D: (0.95, 0.32, -0.053)$.
Whereas for $\theta=(0.1,1)$ the SEH remains in (i), i.e.  in the ultra slow 
ion acoustic regime, for $\theta=10$ (and 30,50) the SEH makes a transition 
from (ii) to (iii), i.e.  it accelerates and jumps from (ii) to (iii) crossing 
the gap of no solution (``forbidden region") to settle in the supersonic 
regime. 

The reason for an additional velocity gain by the solutions in some cases, 
even after achieving the valid set of parameters to qualify as solution of 
the nonlinear Vlasov equation,
 is that the SEH prefers to settle in a region of $\bf{negative}$ free energy
 \cite{GS02}. This enables the 
plasma, by approaching such lower energy state, to gain (harvest) energy which 
during the evolution is deposited for example as heat and/or in other 
fluctuations or excitations. This lower energy state is hence attractive and 
thus preferred by the plasma, which explains the additional acceleration 
observed for large $\theta$ even when the solutions are allowed by the 
NDR \cite{mandal:dnjp}. 

With respect to the pure kinetic effects of ions reflection as treated by 
Dupree 
\cite{dupree83:t} not significantly visible in present cases \cite{mandal:dnjp}, 
the simulations highlight the dominant role of streaming ion populaion in
comparison to the reflected ion population duly accounted for in the 
present simulations. Note that the width, along the velocity dimension, of 
the ion trapping region is smaller by a factor $\sqrt{\theta/\delta}$ in 
comparison to that of electros because of higher kinetic energy of ions 
at similar velocities. In other words, a small amplitude structure would 
not trap/reflect as large fraction of ions density as that of electrons.
Although this small reflected ion population effectively represnts a 
trapped ion population in our periodic setup, it does not
maintain its identity, distinct form the streaming population, over its 
longer transit between two consequitive reflections 
(more so in the limit $L\rightarrow \infty)$.
This justifies neglecting the ion trapping term $b(\alpha,u_{0})$, 
as in the NDR Eq.~(\ref{NDR_eq_Apdx}), since reftected/trapped ions may not 
effectively maintain an $\alpha$ value different from the unity. 
Moreover, for present EH having small $\psi<<T_{e}$ the net momentum transfered, 
because of finite $\partial f_{i}/\partial v$, by the imbalanced populations 
of reflected ions to the hole,
as considered by Dupree \cite{dupree83:t}, is negligible given the
narrow width of ion trapping region along the velocity dimension, as 
discussed above. Therefore, the response of streaming ions remains 
the most dominant factor in determining the observed stability of 
the hole solutions, as considered in the present analysis.

We close this theoretical part with a few experimentally relevant remarks
on spontaneous acceleration of holes.
A similar sudden acceleration of holes (in this case of a periodic train of
ion holes) has been seen in the experiments of \cite{FKPS01}. Density
fluctuation measurements in a double plasma device show an apparently
spontaneous transition of these periodic structures from slow ion acoustic to
ion acoustic velocity regime. In this experiment  gradual scattering of the
trapped ion population by elastic collisions with neutrals was made
responsible for this transition. The outcome of our paper, however, suggests a
second possibility as an alternative explanation, namely the tendency of the
plasma to achieve a lower energy status during the evolution, a process which
will be the more probable the more dilute the plasma is.
\section{Missing hump and extension of the basic EH theory\label{discussion}}
We now indicate an advanced feature of the Electron hole solutions
identified in the present simulation output which might require extension of
the basic EH theory to include a newer parameter to accommodate 
multiplicity of scales in trapped electron density.

Note that the solution in Sec.~\ref{model} are discussed under the 
approximation $k_{0}^{2}=0$, appropriate for a solitary EH with depressed 
trapped electron electron density. 
However, when we examine the 
numerically recovered values of quantity $k$ carefully,
the basic EH theory for these numerical $k$ values prescribes that for 
small $\theta$ solution the electron
density must feature a hump like 
profile. The plasma, however, 
avoids this less stable state by transiting to a multiple-scale state of 
the trapped density where the central phase-space density of the trapped 
electrons still features a sharp dip, surrounded by a relatively flatter 
density profiles. Quantitatively, this situation is resolvable only by 
introducing more sophistication in the hole theory, which is a topic 
addressed in a forthcoming article dedicated to this issue and such a 
generalizing modification of the EH theory \cite{MSS19}.
\section{Summary and Conclusions \label{conclusion}}
In summary, we have proved numerically and theoretically the stable 
existence of hole solutions 
in subcritical plasmas occurring at very low ion temperature values. This 
outcome is striking as it manifests the electron trapping nonlinearity 
as the ruling agent in this evolutionary process standing outside the 
realm of linear wave theory. 
We could show that the potential $\phi(x)$ is essentially a local 
property of the resonant electrons in phase space (via $\beta$) whereas 
the dynamics or hole speed is governed in this slow and ultra slow 
velocity regime by an optimum of electron shielding 
$(-\frac{1}{2}Z_{r}^{'}\left(\tilde{v}_{D}/\sqrt{2}\right)=1)$ and a 
variable, $T_i$ - dependent, ion shielding.
We illustrated that it is the ion response which  
``destabilizes'' the electron hole structure when
$T_{i}<T_{e}/3.5$, causing the slower hole to accelerate, to jump over a forbidden velocity interval
 and to approach a higher speed settling  in the high velocity wing of the electron distribution at a lower energy plasma state. 
For higher ion temperatures $T_{i}>T_{e}/3.5$ these slow holes have already achieved
this status by a reversal of ion shielding that marginalizes the cold ion response. Related to phase-space topology of the trapped electrons in ultra slow EH the present simulation has importantly indicated that in order to model the observed density dip involving trapping with multiple scales, a further modification of the basic EH theory would be necessary. The same would be subject of a forthcoming article \cite{MSS19}. Note that the apprximate analytical threshold, 
$T_{i}^*=T_{e}/3.5$, for existance of coherent solutions may also be subject to modification in any further improved formulation. It is however remarkable 
that the presently obtsained threshold is, at least, of the same order as in the 
simulations since a case with somewhat exagerated value, 
$\theta=10$, is examined in the present simulations for its clarity of results. 

Our description utilizes the  Vlasov equation in its 
full nonlinear version rather than the truncated linear Vlasov 
version. 
%
%
The observed structures and corresponding quantitative analysis 
presented provide the foundation for treating the mechanism of nonlinear 
stability in a number of conditions of high physical relevance where ion 
temperature either approaches or exceeds the electron temperature. 
The explanation of parallel activity in low frequency turbulence phenomena 
in the edge of magnetized fusion plasmas can be further supplemented by
such slow structures. The drift-wave turbulence remains the basic
model for the perpendicular activity in such magnetized conditions.
While stellar or magnetospheric plasmas with nonthermal species are prime 
candidates,
hot plasmas where 
$T_{i}>T_{i}^*$ 
with hole structures caused by a 
reversed ion shielding
may be found in the edge of International Thermonuclear Experiment 
Reactor (ITER) \cite{ITER}, or in the desired operating limit 
of the transport in current free core plasmas of helical confinement 
devices like 
LHD \cite{Okamoto:m} and in modern stellarators like W7-X \cite{dinklage:a}.
 
\appendix
\section{Analytical Model of solitary electron hole (SEH)}
The analytic expression of electron and ion distribution 
function for a solitary electron hole (SEH) solution in presence of electron 
current in a Vlasov Plasma 
system, are given by H. Schamel \cite{schamel:h1} 
\begin{eqnarray}\nonumber
f_{e}(x,v)=\frac{1+K}{\sqrt{2\pi}} 
\begin{cases}
\exp{\left[-\frac {1}{2}\left(\sigma_{e}\sqrt{2\epsilon_{e}}
-\tilde{v}_{D}\right)^2\right]}, &\text{$\epsilon_{e}> 0$}
\\
\exp{\left(-\tilde{v}_{D}^{2}/2\right)}
\exp{\left(-\beta\epsilon_{e}\right)}, &\text{$\epsilon_{e}\leq 0$;}
\end{cases}\\
\label{f-schamel_e}
\end{eqnarray}
\begin{eqnarray}\nonumber
f_{i}(x,u)=\frac{1+A_{e}}{\sqrt{2\pi}} 
\begin{cases}
\exp{\left[-\frac {1}{2}\left(\sigma_{i}\sqrt{2\epsilon_{i}}
+{u}_{0}\right)^2\right]}, &\text{$\epsilon_{i}> 0$}
\\
\exp{\left(-{u}_{0}^{2}/2\right)}
\exp{\left(-\alpha\epsilon_{i}\right)}, &\text{$\epsilon_{i} < 0$;}
\end{cases}\\
\label{f-schamel_i}
\end{eqnarray}
Where, $\epsilon_{e}= v^{2}/2-\phi(x)$, $\epsilon_{i}=u^{2}/2+\theta\left(\phi(x)-\psi\right)$,
$x$ is normalized to $\lambda_{De}=\sqrt{T_{e}/4\pi n_{e}^2}$
and $v$ is normalized by the electron thermal velocity, $v_{the}=
\sqrt{T_{e}/m_{e}}$, $u$ is normalized by ion thermal velocity, 
$v_{thi}=\sqrt{T_{i}/m_{i}}$.The relation between $u$ and $v$ is, $u=\mu v$. 
Where, $\mu=(m_{i}T_{ef}/m_{e}T_{if})^{1/2}$. $v_{0}$ and $u_{0}$ are the phase
 velocity of the wave in the electron and ion frame respectively. 
$\tilde {v}_{D}=v_{D}-v_{0}$, where, $v_{D}$ is the drift velocity of the 
electron. $K=k_{0}^{2}\psi/2$ is the wave number.
Variations with $k_{0}^2>0$ and $k_{0}^2=0$ correspond to a sinusoidal wave and 
a solitary wave solution, respectively. $A_{e}$ is also a constant. 
$\theta=T_{ef}/T_{if}$. $\alpha$ and $\beta$ are the trapping parameter. 
Velocity integration of the Eq.~(\ref{f-schamel_e}) and (\ref{f-schamel_i}) 
yields in small amplitude limit i.e, $\psi \ll 1$, 
\begin{eqnarray}
n_{e}(\phi) \approx 1-\frac{1}{2}Z_{r}^{'}(\tilde{v}_{D}/\sqrt{2})\phi(x)
-\frac{4}{3}b(\beta,\tilde{v}_{D})\phi(x)^{3/2}+..
\vphantom{\frac{1}{2}},
\label{density_e_Apdx}
\end{eqnarray}
\begin{eqnarray}\nonumber
n_{i}(\phi)&\approx&1-\frac{1}{2}Z_{r}^{'}(u_{0}/\sqrt{2})
\theta(\psi-\phi(x)) \\
&&-\frac{4}{3}b(\alpha,u_{0})\vphantom{\frac{1}{2}}
\left[\theta(\psi-\phi(x))\vphantom{\frac{1}{2}}\right]^{3/2}+...,
\label{density_i_Apdx}
\end{eqnarray}
where $b(\beta,\tilde{v}_{D})$ and $b(\alpha,u_{0})$
determine the trapped particle density of electron and ion, respectively.
In absence of ion trapping $b(\alpha,u_{0})=0$ and $b(\beta,\tilde{v}_{D})$ is
 given by:
\begin{eqnarray}\nonumber
b(\beta,\tilde{v}_{D})=\frac{1}{\sqrt{\pi}}\left(1-\beta-u_{0}^{2}\right)
\exp(-\tilde{v}_{D}^2/2)
\label{b_beta}
\end{eqnarray}
In the Eq.~(\ref{density_e_Apdx}) and (\ref{density_i_Apdx})
$Z_{r}(x)=-2e^{-x^{2}}\int_{0}^{x}dt\exp{(t^2)}$
is the real part of the plasma dispersion function and $\phi$ is the potential,
 satisfying the Poisson equation.
\begin{eqnarray}
\frac{\partial^{2}\phi}{\partial x^{2}}=n_{e}(\phi)-n_{i}(\phi)\equiv 
-\frac{\partial V(\phi)}{\partial \phi}
\label{Poisson_eq}
\end{eqnarray}
The Sagdeev pseudo potential $V(\phi)$ associated with the Poisson equation 
Eq.~(\ref{Poisson_eq}) is given by:
\begin{eqnarray}
-V(\phi)=\frac{k_{0}^{2}}{2}\phi(\psi-\phi)+\frac{8}{15}
b(\beta,\tilde{v}_{D})
\phi^{2}\left(\sqrt{\psi}-\sqrt{\phi}\right)
\label{pseudo_pot_Apdx}
\end{eqnarray}
The phase velocity of the structure is determined through the Nonlinear 
Dispersion Relation (NDR) Eq.~(\ref{NDR_eq_Apdx}) in terms of $\alpha$, $\beta$ 
and $k_{0}^2$ 
\begin{eqnarray}\nonumber
k_{0}^2&-&\frac{1}{2}Z_{r}^{'}\left(\tilde{v}_{D}/\sqrt{2}\right)
-\frac{\theta}{2}Z_{r}^{'}\left(u_{0}/\sqrt{2}\right)\\
&=&\frac{16}{15}b(\beta,\tilde{v}_{D})\psi^{1/2} = B
\label{NDR_eq_Apdx}
\end{eqnarray}
We define $B := \frac{16}{15}b(\beta,\tilde{v}_{D})\psi^{1/2}$. 
Substituting, Eq.~(\ref{pseudo_pot_Apdx}) in Eq,~(\ref{Poisson_eq}) and 
subsequent integration in the limit $k_{0}^2\rightarrow 0$, applicable to the 
existence of the solitary solution,
leads to the following solutions for the potential structure 
$\phi(x)$ in terms of the parameters coming from NDR Eq.~(\ref{NDR_eq_Apdx}):
\begin{eqnarray}\nonumber
\phi(x)=\psi \sech^{4}\left(\sqrt{\frac{b(\beta,\tilde{v}_{D})}{15} 
\sqrt{\psi}}x\right)\\
= \psi \sech^4\left({\frac{\sqrt{B}x}{4}}\right)
\label{soliton}
\end{eqnarray}  
which requires a positive $B$, $B>0$.

\end{document}